\documentclass[12pt,a4paper]{article}
\usepackage{mathrsfs}
\usepackage{epsfig}
\begin{document}
\title{ Photoproduction of the charged top-pions at the LHeC}
\author{Chong-Xing Yue,  Jing Guo, Jiao Zhang, Qing-Guo Zeng \\
{\small Department of Physics, Liaoning  Normal University, Dalian
116029, P. R. China}
\thanks{E-mail:cxyue@lnnu.edu.cn}}
\date{\today}

\maketitle
\begin{abstract}
The top triangle moose $(TTM)$ model, which can be seen as the
deconstructed version of the topcolor-assisted technicolor ($TC2$)
model, predicts the existence of the charged top-pions $
\pi_{t}^{\pm}$ in low energy spectrum. In the context of this model,
we consider photoproduction of $\pi^{\pm}_{t}$ via the subprocesses
$\gamma b\rightarrow t \pi_{t}^{-}$ and $\gamma
\overline{b}\rightarrow \overline{t} \pi_{t}^{+}$ at the large
hadron-electron collider ($LHeC$), in which high energy photon beams
are generated by using the Compton backscatting method. We find
that, as long as the charged top-pions are not too heavy, they can be
abundantly produced via $\gamma b$ collision.

\vspace{0.5cm} \vspace{2.0cm} \noindent
 {\bf PACS numbers}: 12.60.Cn, 14.80.Cp, 13.85.Rm

\end{abstract}
\newpage
\noindent{\bf 1. Introduction }\vspace{0.5cm}

Although the standard model $(SM)$ is an excellent low energy effective
theory, the cause of electroweak symmetry breaking $(EWSB)$ and the
origin of fermion masses continue to be outstanding mystery, which has started to be probed at the $LHC$. Recently,
the $ATLAS$ and $CMS$ collaborations have found some hints of a relatively light Higgs boson with a mass somewhere
between 120 and 130 $GeV$ [1]. Thus we are now coming into an exciting period of particle
physics.

 The top quark is the heaviest elementary particle known
up to today. Its mass might has a different origin from the masses of
other quarks and leptons, a top quark condensate, $<t\bar{t}>$,
could be responsible for at least part of $EWSB$. Much theoretical
work has been carried out in connection to the top quark and $EWSB$,
some specific new physics models are proposed (for review, see [2]
and references therein). An interesting model involving a role for the top quark in
dynamical $EWSB$ is known as the topcolor-assisted technicolor ($TC2$) model [3]. In the
framework of this model, the topcolor, a new $QCD-like$ interaction,
that couples strongly to the third generation quarks, makes small
contributions to $EWSB$ and gives rise to the main part of the top
quark mass. Technicolor $(TC)$ provides the bulk of $EWSB$ via the
vacuum expectation value $(VEV)$ of a technifermion bilinear. The
light fermions can get masses from the extended technicolor $(ETC)$.
The $TC2$ model is one of well-motivated new physics models and has all essential
features of the topcolor scenario.

Higgsless models [4] have emerged as a novel way of understanding
the mechanism of $EWSB$ without the presence of a scalar particle in
the spectrum. Recently, combing Higgsless and topcolor mechanisms, a
deconstructed Higgsless model was proposed, called the top triangle
moose $(TTM)$ model [5, 6]. In this model, $EWSB$ results largely
from the Higgsless mechanism while the top quark mass is mainly
generated by the topcolor mechanism. The $TTM$ model alleviates the
tension between obtaining the correct top quark mass and keeping
$\Delta\rho$ small that exists in many Higgsless models, which can
be seen as the deconstructed version of the $TC2$ model.

The new physics models belonging to the topcolor scenario
genetically have two sources of $EWSB$ and there are two sets of
Goldstone bosons. One set is eaten by the electroweak  gauge bosons
$W$ and $Z$ to generate their masses, while the other set remans in
the spectrum, which is called the top-pions ($\pi_{t}^{0}$  and
$\pi^{\pm}_{t}$). Topcolor scenario also predicts the existence of
the top-Higgs $h^{0}_{t}$, which is the $t\bar{t}$ bound state. It
is well known that the possible signals of these new scalar
particles have been extensively studied in the literature. However,
most of works are done in the context of the $TC2$ model.

More phenomenology analysis about the top-pions and top-Higgs predicted
by the $TTM$ model is needed. Although photoproduction
of the charged top-pions has been studied in Ref.[7], which proceeds
via the subprocess $\gamma c\rightarrow b \pi_{t}^{+}$ mediated by
the flavor changing couplings. The high energy photon beams are generated
 by using the Compton backscatting of the initial electron and laser photon beams.
 However, so far, photoproduction of
the charged top-pions via the subprocesses $\gamma b\rightarrow t
\pi_{t}^{-}$ and $\gamma \overline{b}\rightarrow \overline{t}
\pi_{t}^{+}$ has not been considered at the large hadron-electron
collider ($ LHeC$).  Furthermore, many popular models beyond the $SM$ predict the existence of the charged scalars. At the $LHeC$,
these new particles can be produced via $\gamma b$ collision, which has not been
 detailed studied in the literature, as we know.  Thus, in this paper, we will consider
photoproduction of the charged top-pion associated  with a top quark
in the frameworks of the $TTM$ and $TC2$ models and compare the
numerical results with each other. Our calculation can be easily transformed to other models. We hope that our works will be
helpful to test topcolor models and further to distinguish different new physics models at the $LHeC$.

The layout of the present paper is as follows. After reviewing the
essential features of the $TTM$ model in section 2, we calculate the
production cross section of the subprocess $\gamma b\rightarrow t
\pi_{t}^{-}$ at the $LHeC$ in section 3. To compare our numerical
results with those of the $TC2$ model, we further consider
photoproduction of the charged top-pions predicted by the $TC2$ model
in section 4. Our conclusion and discussion are given in section 5.

\vspace{0.5cm} \noindent{\bf 2. The essential features of the $TTM$
model }

\vspace{0.5cm}The detailed description of the $TTM $ model can be
found in Refs.[5, 6], and here we just want to briefly review its
essential features, which are related to our calculation.

The electroweak gauge structure of the $TTM$ model is
$SU(2)_{0}\times SU(2)_{1}\times U(1)_{2}$. The nonlinear sigma
field $\sum_{01}$ breaks the group $SU(2)_{0}\times SU(2)_{1}$ down to
$SU(2)$ and field $\sum_{12}$ breaks $SU(2)_{1}\times U(1)_{2}$
down to $U(1)$. To separate top quark mass generation from $EWSB$, a
top-Higgs field $\Phi$ is introduced to the $TTM$ model, which
couples preferentially to the top quark. To ensure that most of the $EWSB$
comes from the Higgsless side, the $VEVs$ of the fields $\sum_{01}$
and $\sum_{12}$ are chosen to be $<\sum_{01}>=<\sum_{12}>=F=\sqrt2
\nu\cos\omega$, in which $\nu=246 GeV$ is the electroweak scale and
$\omega$ is a new small parameter. The $VEV$ of the top-Higgs field
is $f=<\Phi>=\nu \sin\omega$.

From above discussions, we can see that, for the $TTM$ model, there
are six scalar degrees of freedom on the Higgsless sector and four
on the top-Higgs sector. Six of these Goldstone bosons are eaten to
give masses to the gauge bosons $W^{\pm}$, $Z$, $W'^{\pm}$ and $Z'$.
Others remain as physical states in the spectrum, which are called
the top-pions ($\pi_{t}^{\pm}$ and $\pi_{t}^{0}$) and the top-Higgs
$h_{t}^{0}$. In this paper, we will focus our attention on
photoproduction of the charged top-pions at the $LHeC$. The
couplings of the charged top-pions $\pi_{t}^{\pm}$ to ordinary
particles, which are related our calculation, are given by [6]
\begin{eqnarray}
{\cal
L}_{\pi_{t}tb}=i\lambda_{t}\cos\omega\{1-\frac{x^{2}[a^{4}+(a^{4}-2a^{2}+2)\cos2\omega]}{8(a^{2}-1)^{2}}\}
(\pi_{t}^{+}\bar{t}bP_{L}+\pi_{t}^{-}t\bar{b}P_{R})
\end{eqnarray}
with
\begin{eqnarray}
\lambda_{t}=\frac{\sqrt{2}m_{t}}{\nu\sin\omega}[\frac{M_{D}^{2}(\varepsilon_{L}^{2}+1)
-m_{t}^{2}}{M_{D}^{2}-m_{t}^{2}}],
\hspace{0.5cm} a=\frac{\nu\sin\omega}{\sqrt{2}M_{D}},\hspace{0.5cm}
x=\sqrt{2}\varepsilon_{L}=\frac{2\cos\omega M_{W}}{M_{W'}}.
\end{eqnarray}
Here $P_{L(R)}=\frac{1}{2}(1\mp\gamma_{5})$ is the
left(right)-handed projection operator, $M_{D}$ is the mass scale of
the heavy fermion and $M_{W'}$ is the mass of the new gauge boson
$W'$. Since the top quark mass depends very little on the
right-handed delocalization parameter $\varepsilon_{tR}$, we have
set $\varepsilon_{tR}=0$ in $Eq.(1)$. The parameter
$\varepsilon_{L}$ describes the degree of delocalization of the
left-handed fermions and is flavor universal, the parameter $x$
presents the ratio of gauge couplings. The relationship between
$\varepsilon_{L}$ and $x$, which is given in  $Eq.(2)$, is imposed by ideal delocalization.

Reference [8] has shown that $M_{W'}$ should be larger than $380GeV$
demanded by the $LEPII$ data and smaller than $1.2TeV$ by the
need to maintain  perturbative unitarity in $W_{L} W_{L}$ scattering.
It is obvious that the coupling $\pi_{t}tb$ is not very sensitive to
the parameters $M_{W'}$ and $M_{D}$. Thus, the production cross sections of the subprocesses
$\gamma b\rightarrow t \pi_{t}^{-}$ and $\gamma \overline{b}\rightarrow \overline{t}
\pi_{t}^{+}$ are not strongly dependent on the values of the mass parameters $M_{W'}$ and $M_{D}$.
In our following numerical
calculation, we will take the illustrative values $M_{W'}=500GeV$
and $M_{D}=650GeV$. In this case, there is
$[M_{D}^{2}(\varepsilon_{L}^{2}+1)-m_{t}^{2}]/(M_{D}^{2}-m_{t}^{2})\approx1$
and Eq.(1) can be approximately written as
\begin{eqnarray}
{\cal L}_{\pi_{t}tb}\approx i\frac{\sqrt{2}m_{t}C}{\nu}\cot\omega
(\pi_{t}^{+}\bar{t}bP_{L}+\pi_{t}^{-}t\bar{b}P_{R})
\end{eqnarray}
with
\begin{eqnarray}
C=1-\frac{x^{2}[a^{4}+(a^{4}-2a^{2}+2)\cos2\omega]}{8(a^{2}-1)^{2}}.
\end{eqnarray}
It is obvious that constant $C$ is not sensitive to the value of $\sin\omega$ and its value close to 1.

\vspace{0.5cm} \noindent{\bf 3. Photoproduction of the charged
top-pions at the $LHeC$ within the $TTM$ \hspace*{0.6cm} model }

\vspace{0.5cm} Recently, the high-energy $ep$\hspace{0.1cm}
collision has been considered at the $LHC$, which is called the
$LHeC$ [9, 10]. At the $LHeC$, the incoming proton beam has an energy
$E_{p}=7TeV$ and the energy $E_{e}$ of the incoming electron is in the
range of $50\sim200GeV$, corresponding to the center-of-mass
$(c.m.)$ energy of $\sqrt{s}=2\sqrt{E_{p}E_{e}}\approx
1.18\sim2.37TeV$. Its anticipated integrated luminosity is at the
order of $10\sim100fb^{-1}$ depending on the energy of the incoming
electron and the design. The $LHeC$ can be used to accurately
determine the parton dynamics and the momentum distributions of
quarks and gluons in proton. Furthermore, it can provide better
condition for studying a lot of phenomena comparing to the high
energy linear $e^{+}e^{-}$ collider $(ILC)$ due to the high $c.m.$
energy and to the $LHC$ due to more clear environment $[10, 11]$.
Thus, it may play a significant role in the discovery of new physics
beyond the $SM$. An other advantage of the $LHeC$ is the opportunity
to construct $\gamma p$ collider $[12]$ with the photon beam generated by the backward
Compton scattering of incident electron- and laser-beams. The energy
and luminosity of the photon beam would be the same order of
magnitude of the parent electron beam.

\begin{figure}[htb]
\vspace{0.5cm}
\begin{center}
 \epsfig{file=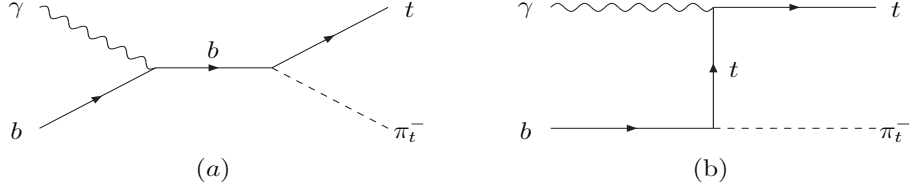,width=340pt,height=70pt}
\vspace{0.2cm} \caption{ Feynman diagrams for the subprocess $\gamma
b\rightarrow t\pi_{t}^{-}.$} \label{ee}
\end{center}
\end{figure}

From the discussions given in section 2, we can see that the charged
top-pion $\pi_{t}^{-}$ can be produced via the subprocess $\gamma(k_{1})
b(p_{1})\rightarrow t(p_{2})\pi_{t}^{-}(k_{2})$ at the $LHeC$. The relevant Feynman
diagrams are depicted in $Fig.1$. With the relevant couplings, the
invariant production amplitude can be written as
\begin{eqnarray}
M\!\!\!\!\!\!&&=M_{a}+M_{b}
  \nonumber\\&&=A\hspace{0.1cm}\bar{u}(p_{2})\hspace{0.1cm}P_{R}\frac{i}{\not\!\!\hspace*{0.12cm}k_{1}
  +\not\!\!\hspace*{0.12cm}p_{1}}
  (-\frac{1}{3}e\gamma^{\mu})\hspace{0.15cm}u(p_{1})\hspace{0.15cm}\varepsilon_{\mu}(k_{1})
\nonumber\\&&\hspace*{0.3cm}+A\hspace{0.1cm}\bar{u}(p_{2})\hspace{0.1cm}(\frac{2}{3}e\gamma^{\mu})
\frac{i}{\not\!\!\hspace*{0.12cm}p_{2}-\not\!\!\hspace*{0.12cm}k_{1}-m_{t}}
P_{R}\hspace{0.15cm}u(p_{1})\hspace{0.15cm}\varepsilon_{\mu}(k_{1})
\end{eqnarray}
with
\begin{eqnarray}
A=\frac{\sqrt{2}m_{t}C}{\nu}\cot\omega.
\end{eqnarray}
Here $\varepsilon_{\mu}(k_{1})$ is the polarization vector of the
photon. In above equation,\hspace{0.15cm}we have taken $m_{b}\approx
0$. Using the amplitude $M$, we can directly obtain the cross
section $\hat{\sigma}(\hat{s})$ for the subprocess $\gamma
b\rightarrow t\pi_{t}^{-}$.

Since the photon beam in $\gamma b$ collision is generated by the Compton backscattering of the incident electron- and the laser-beams,
the effective cross section $\sigma_{1}(s)$ at the $LHeC$ can be
obtained by folding $\hat{\sigma}(\hat{s})$ with the bottom quark
and photon distribution functions. For this purpose, we define the
variables: $\hat{s}=x_{1}x_{2}s$ with $x_{1}=E_{\gamma}/E_{e}$ and
$x_{2}=E_{b}/E_{p}$.
 $\sqrt{s}$ is the $c.m.$ energy of the $LHeC$.  The cross section
 $\sigma_{1}(s)$ for the process
$ep\rightarrow \gamma b+X\rightarrow t\pi_{t}^{-}+X$ can be given by
\begin{eqnarray}
\sigma_{1}(s)=\int^{x_{1max}}_{(m_{\pi_{t}}+m_{t})^{2}/s}dx_{1}
\int^{1}_{(m_{\pi_{t}}+m_{t})^{2}/x_{1}s}dx_{2}
f_{\gamma/e}(x_{1})f_{b/p}(x_{2})\hat{\sigma}(\hat{s}).
\end{eqnarray}
For unpolarized initial electron and laser beams, the energy spectrum
of the backscattered photon is [13]
\begin{eqnarray}
f_{\gamma/e}(x)=\frac{1}{D(\xi)}\{1-x+\frac{1}{1-x}[1-\frac{4x}{\xi}(1-\frac{x}{\xi(1-x)})]\}
\end{eqnarray}
with
\begin{eqnarray}
D(\xi)=(1-\frac{4}{\xi}-\frac{8}{\xi^{2}})\ln(1+\xi)+\frac{1}{2}+\frac{8}{\xi}-\frac{1}{2(1+\xi)^{2}}.
\end{eqnarray}
Where $\xi=4E_{e}E_{0}/m^{2}_{e}$ in which $m_{e}$ denotes the incident electron mass,
$E_{0}$ denotes the initial laser photon energy. $x$ is the fraction of energy taken by
the backscattered photon beam moving along the initial electron direction.
$f_{\gamma/e}(x)$ vanishes for $x>x_{max}=E_{max}/E_{e}=\xi/(1+\xi)$.
In order to get ride of the background
effects in the Compton backscattering, particularly $e^{+}e^{-}$
pair production in the collision of the laser with the backscattered
photon, it is required $E_{0}x_{max}\leq m_{e}^{2}/E_{e}$ which implies
$\xi\leq2+2\sqrt{2}\approx4.83$ [13]. For the choice $\xi=4.8$, one can obtain
$x_{max}\approx0.83$ and $D(\xi)\approx1.84$.
In $Eq.(7)$, the  bottom quark is directly taken from the proton in a five flavor scheme.
For the bottom quark distribution function $f_{b/p}(x_{2})$,
we will use the form given by the $CTEQ6L$ [14] parton distribution
functions $(PDFs)$. The renormalization and factorization scales are
taken as the $t\pi_{t}^{-}$ invariant mass.

Except for the $SM$ input parameters $\alpha_{e}=1/128$,
$m_{t}=172GeV$, and $M_{W}=80.4GeV$ [14], the production cross
section for the process $ep\rightarrow t\pi_{t}^{-}+X$ is dependent
on the free parameters $\sin\omega$ and $m_{\pi_{t}}$. The parameter $\sin\omega$ indicates
the fraction of $EWSB$ provided by the top condensate. The top-pion mass $m_{\pi_{t}}$ depend on the amount of top-quark mass arising
from the $ETC$ sector and on the effects of electroweak gauge interactions [2], and thus its value is model-dependent. In the context of the $TTM$ model, Ref.[6] has obtained  the constraints on the top-pion mass via studying its effects on the relevant experimental observables. Similarly with
Refs.[6, 16], we will assume that the values of the free parameters $\sin\omega$ and $m_{\pi_{t}}$ are in the ranges of
$0.2 \sim 0.8$ and $200 \sim 600GeV$, respectively.

\begin{figure}
{\includegraphics[scale=0.68]{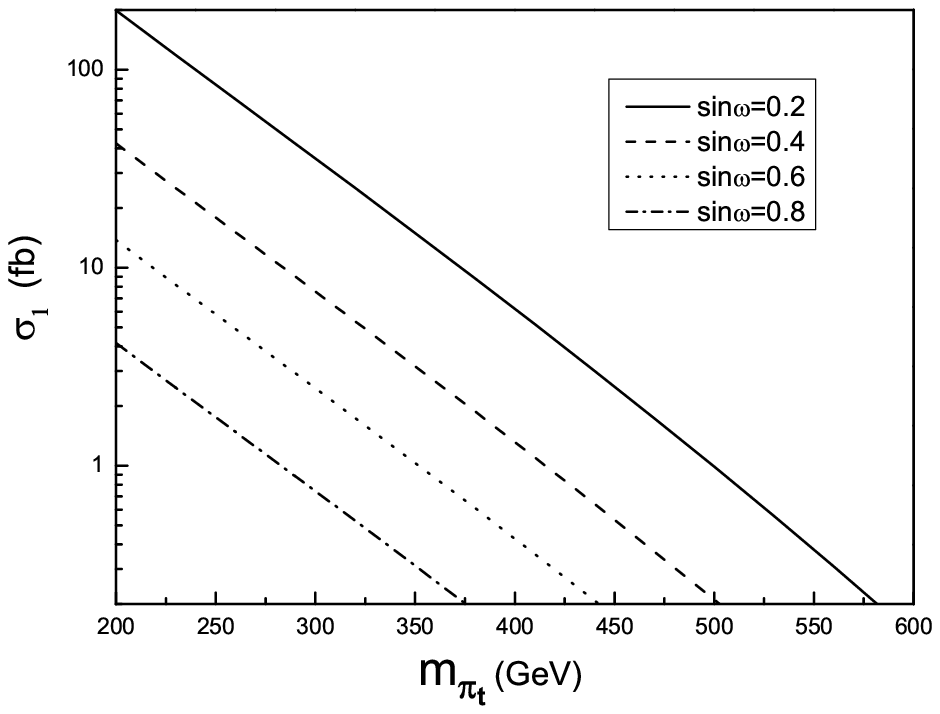}}
{\includegraphics[scale=0.68]{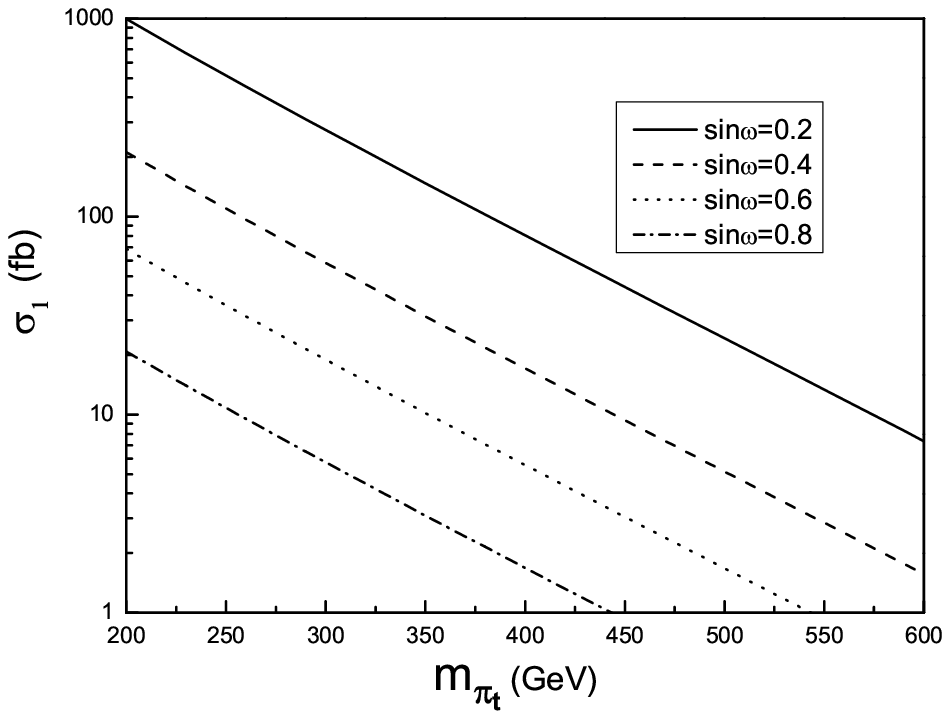}}
{\includegraphics[scale=0.68]{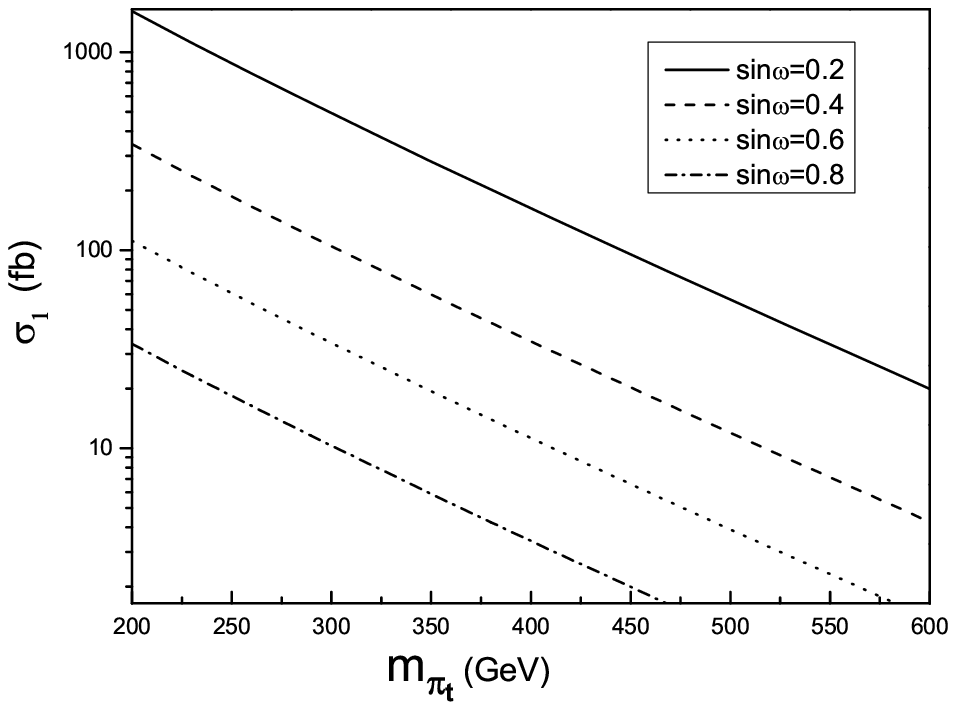}}
\caption{The cross section $\sigma_{1}(s)$ as a function of the
 mass parameter $m_{\pi_{t}}$ for $E_{e}=\hspace*{1.7cm} 70GeV$ (a), $150GeV$ (b) and
$200GeV$ (c).}\label{ee}
\end{figure}

Our numerical results are summarized in $Fig.2$, in which we have
plotted the production cross section $\sigma_{1}(s)$ as a function of
the charged top-pion mass $m_{\pi_{t}}$ for $E_{e}=70GeV$, $150GeV$
and $200GeV$, and various values of the parameter $\sin\omega$. One
can see that, for $70GeV\leq E_{e}\leq 200GeV$, $200GeV\leq
m_{\pi_{t}}\leq 600GeV$, and $0.2\leq \sin\omega \leq0.8$, the value
of the cross section $\sigma_{1}(s)$ for the process $ep\rightarrow
t\pi_{t}^{-}+X$ is in the range of $2.8\times 10^{-3}fb\sim
1.6\times 10^{3}fb$, which is sensitive to the free parameters
$\sin\omega$ and $m_{\pi_{t}}$. If we assume that the yearly
integrated luminosity ${\cal L}_{Lint}=50fb^{-1}$, than there will
be several and up to ten thousands of $t\pi_{t}^{-}$ events to be
generated at the $LHeC$ per year.

It is well known that, for a heavy charged scalar, which is heavier
than the top quark, the main production channel proceeds via gluon-bottom
collision at the $LHC$. In the context of the $TTM$ model, Ref.[6] has
studied production of the charged top-pion $\pi_{t}^{-}$ via the subprocess
$gb\rightarrow t\pi_{t}^{-}$ and discussed the possibility of detecting the
charged top-pions at the $LHC$. It is obvious that the $t\pi_{t}^{-}$
production cross section at the $LHC$ with $\sqrt{s}=14TeV$ is larger
than that at the $LHeC$ with $\sqrt{s}=2.37TeV$ as shown in Fig.2. However,
considering the more clear environment comparing to the $LHC$,
it is needed to consider the subprocess $\gamma b\rightarrow t\pi_{t}^{-}$
in the near future $LHeC$ experiments.
\begin{figure}[htb]
\vspace{-0.5cm}
\begin{center}
 \epsfig{file=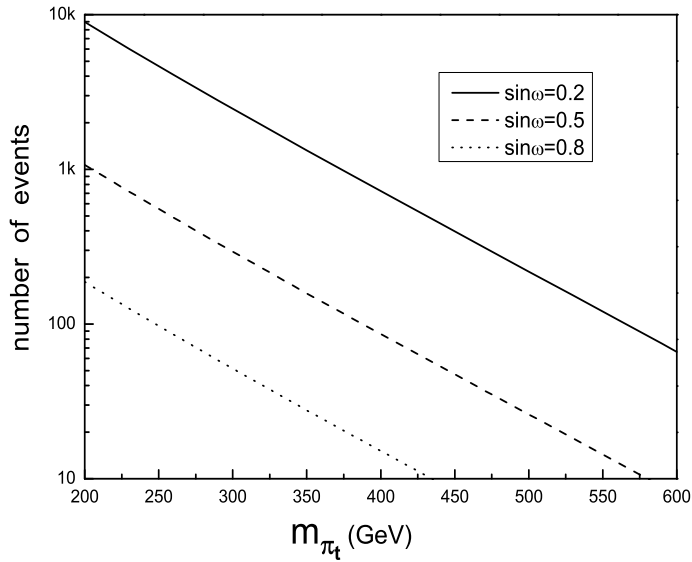,width=300pt,height=260pt}
 \vspace{-0.5cm}
 \caption{The number of the $l^{+}l^{-}+bbb+\;
{\not\!\!E}\;$ signal
events generated at the $LHeC$.} \label{ee}
\end{center}
\end{figure}

Reference [6] has shown that, for $m_{h_{t}}\geq 300GeV$ and
$m_{\pi_{t}}\leq 600GeV$, the charged top-pions $\pi_{t}^{\pm}$
dominantly decay into $tb$ and there is
$Br(\pi_{t}^{\pm}\rightarrow tb)>90\%$. Thus,
photoproduction of the charged top-pion associated a top quark can
easily transfer to the $t\bar{t}b$ final state. In order to ensure
the clearest event signature, only fully leptonic decay modes of the
gauge boson $W$ are considered. Then, photoproduction of the
charged top-pions can give rise to the $l^{+}l^{-}+bbb+\;
{\not\!\!E}\;$ signature at the $LHeC$. Its production rate can be
easily estimated by multiplying the overall decay branching ratios
to the effective production cross section, which can be
approximately written
as: $[\sigma_{1}(t\pi_{t}^{-})+\sigma_{1}(\bar{t}\pi_{t}^{+})]\times[Br(t\rightarrow
Wb)]^{2}\times Br(\pi_{t}^{\pm}\rightarrow
tb)\times[Br(W\rightarrow l\nu)]^{2}\approx 2\sigma_{1}\times
1\times0.9\times(3\times0.108)^{2}\approx0.18\sigma_{1}$. In this estimation,
we have assumed that the production cross section of the subprocess
$\gamma\bar{b}\rightarrow \bar{t}\pi_{t}^{+}$ equals to that for the
subprocess $\gamma b\rightarrow t\pi_{t}^{-}$, and taken
$Br(t\rightarrow Wb)\approx1$ and $Br(W^{\pm}\rightarrow e\nu_{e}) \simeq Br(W^{\pm}\rightarrow \mu \nu_{\mu})\simeq Br(W^{\pm}\rightarrow \tau \nu_{\tau})\simeq 10.8\%$. The number of the signal
events generated at the $LHeC$ per year are given in $Fig.3$, in which
we have taken $E_{e}=150GeV$ and the yearly integrated luminosity
${\cal{L}}_{Lint}=50fb^{-1}$. One can see from this figure that, in
wide range of the parameter space of the $TTM$ model, there will be thousands of the
$l^{+}l^{-}+bbb+\; {\not\!\!E}\;$ signal events to
be generated at the $LHeC$. If the electroweak gauge boson $W$ decays to $l\nu$ with $l$ denoting $e$ or $\mu$, the number of the signal events will be reduced. However, this allows the invariant mass of the charged
top-pion to be reconstructed, which help separate the signal from the
large $t\bar{t}+jets$ background.
 Certainly, detailed confirmation of the
observability of the signals generated by the process $ep\rightarrow
t\pi_{t}^{\pm}+X$, would require Monte Carlo simulation of the
signals and the relevant $SM$ backgrounds, which is beyond the scope
of this paper.

\vspace{0.5cm} \noindent{\bf 4. Photoproduction of the charged
top-pions at the $LHeC$ within the $TC2$ \hspace*{0.6cm} model }

\vspace{0.5cm} To solve the phenomenological difficulties of
traditional $TC$ theory, topcolor models [2] were proposed by combining
$TC$ interactions with the topcolor interactions at the scale of
about $1TeV$. It is well known that the $TC2$ model [3] is one of
the phenomenologically viable models, which has almost all essential
features of this kind of new physics models.

For the $TC2$ model [3], $TC$ interaction plays the main role in
$EWSB$. Topcolor interaction makes small contributions to $EWSB$ and
gives rise to the main part of the top quark mass,
$(1-\varepsilon)m_{t}$, with the parameter $\varepsilon\ll 1$. Thus,
there is the relation
\begin{eqnarray}
\nu_{\pi}^{2}+F_{t}^{2}=\nu_{W}^{2}.
\end{eqnarray}
Where $\nu_{\pi}$ represents the contributions of $TC$
interactions to $EWSB$, $\nu_{W}=\nu/\sqrt{2}=174GeV$.
Here $F_{t}$ is the physical top-pion decay constant,
which can be estimated from the Pagels-Stokar  formula and written as
\begin{eqnarray}
F_{t}^{2}=\frac{N_{c}}{16\pi^{2}}m_{t,dyn }^{2}\ln(\frac{\Lambda^{2}}{m_{t,dyn}^{2}}).
\end{eqnarray}
Where $N_{c}=3$ is the color factor, $\Lambda$ is the cutoff scale and $m_{t,dyn}$
denotes the portion of the top quark mass generated by the topcolor interaction.
In the case of $m_{t,dyn}\approx m_{t}$ and $1TeV\leq\Lambda\leq 20TeV$, Ref.[16]
has shown that the value of the factor $\sin\omega=F_{t}/\nu_{W}$ is in the
range of $0.25$ and $0.5$. Allowing $F_{t}$ to vary over this ranges does not qualitatively
change our conclusion, thus, we will take $F_{t}=50GeV$ for illustration in our numerical analysis,
which corresponds to $\Lambda=1.59TeV$.

In the $TC2$ model, topcolor interaction is not flavor-universal and
mainly couples to the third generation quarks. Thus, the top-pions ($\pi_{t}^{\pm},\pi_{t}^{0}$)
have large Yukawa couplings to the third family.
The explicit forms for the couplings of $\pi_{t}^{\pm}$ to the third
generation quarks, which are related our calculation, can be written
as [3, 17]
\begin{eqnarray}
\frac{(1-\varepsilon)m_{t}}{F_{t}}\frac{\sqrt{\nu_{w}^{2}-F_{t}^{2}}}{\nu_{w}}(\pi_{t}^{+}\bar{t}b P_{L}+\pi_{t}^{-} t \bar{b}P_{R}).
\end{eqnarray}
Where the factor $\sqrt{\nu_{w}^{2}-F_{t}^{2}}/\nu_{w}$ reflects the
effects of the mixing between the top-pion and the electroweak Goldstone boson of the $TC$ sector.

\begin{figure}[htb]
\vspace{-0.5cm}
\begin{center}
 \epsfig{file=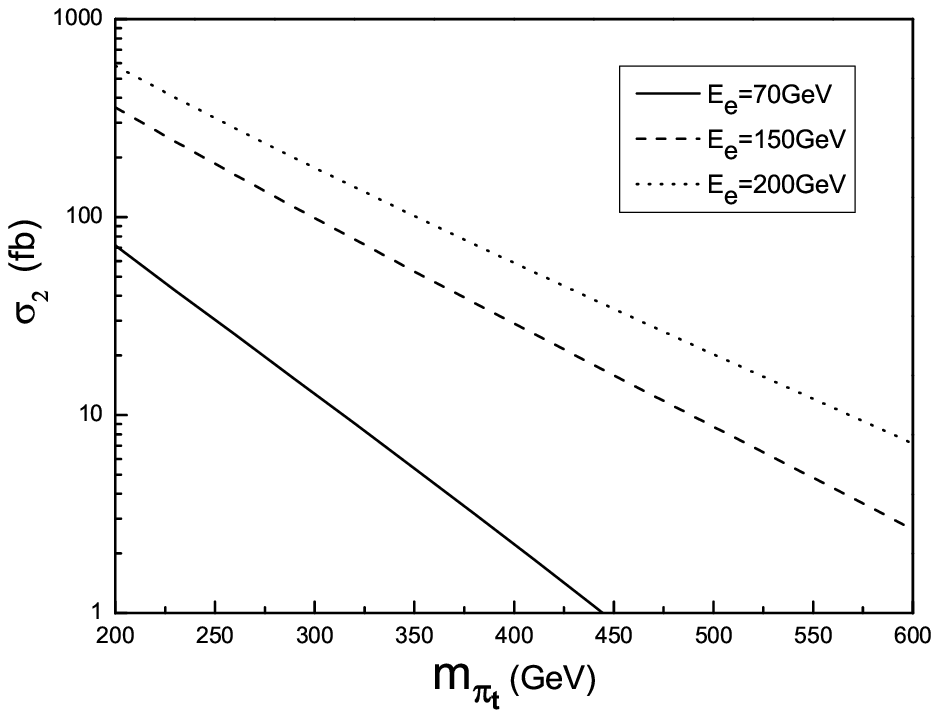,width=320pt,height=280pt}
 \vspace{-0.5cm}
 \caption{The cross section $\sigma_{2}(s)$ as a function of the
 mass parameter $m_{\pi_{t}}$ for \hspace*{2cm} $E_{e}=70GeV$, $150GeV$ and
$200GeV$.} \label{ee}
\end{center}
\end{figure}

It is obvious that the charged top-pions $\pi_{t}^{\pm}$ predicted
the $TC2$ model can also be produced via the subprocesses $\gamma
b\rightarrow t\pi_{t}^{-}$ and $\gamma \bar{b}\rightarrow
\bar{t}\pi_{t}^{+}$ at the $LHeC$. The effective production cross
section $\sigma_{2}$ only depends on two free parameters
$\varepsilon$ and $m_{\pi_{t}}$. The parameter $\varepsilon$
parameterizes the portion of the $ETC$ contributions to the top
quark mass. From theoretical point of view, $\varepsilon$ with value
from $0.01$ to $0.1$ is favored. The  cross section $\sigma_{2}$ depends on the free parameter
$\varepsilon$ only via the factor $(1-\varepsilon)^2$, thus its
value is not sensitive to the free parameter
$\varepsilon$ and we will take $\varepsilon =0.05$ in our numerical estimation. For the top-pion
mass, the mass splitting between the neutral top-pion and the
charged top-pion is very small, since it comes only from the
electroweak interactions [18]. The absence of $t\rightarrow
\pi_{t}^{+}b$ implies that $m_{\pi_{t}^{+}}>165GeV$ [19] and the
branch ratio $R_{b}$ for the decay $Z\rightarrow b\bar{b}$ yields
$m_{\pi_{t}^{+}}>220GeV$ [20]. In this paper, we will take
$m_{\pi_{t}^{0}}=m_{\pi_{t}^{\pm}}=m_{\pi_{t}}$ and assume that its value is in
the range of $200\sim600GeV$.
\begin{figure}[htb]
\vspace{-0.5cm}
\begin{center}
 \epsfig{file=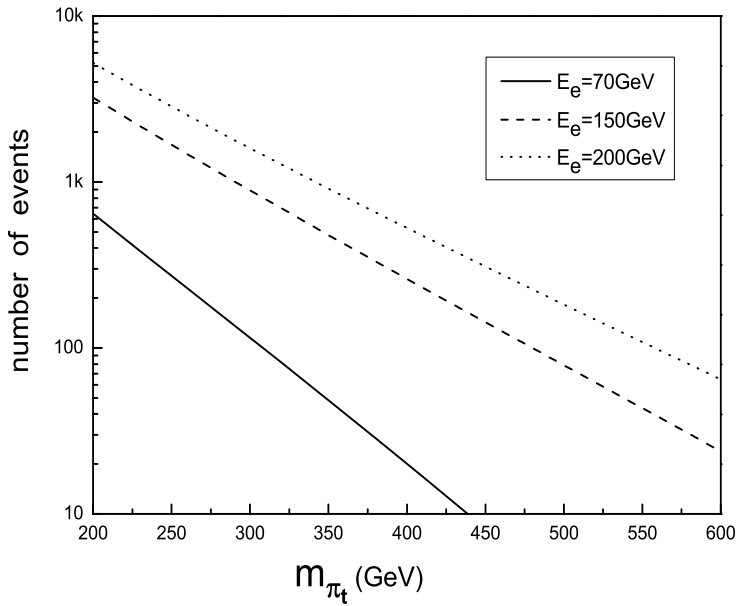,width=320pt,height=280pt}
 \vspace{-0.5cm}
 \caption{The number of the
$l^{\pm}+bb+jet+\; {\not\!\!E}\;$ events generated at the $LHeC$.} \label{ee}
\end{center}
\end{figure}

The production cross section $\sigma_{2}$ for the process
$ep\rightarrow t\pi_{t}^{-}+X$ is plotted in $Fig.4$ as a function
of the mass parameter $m_{\pi_{t}}$ for the  parameters
$\varepsilon=0.05$ and $E_{e}=70GeV$, $150GeV$, $200GeV$. One can see
from this figure that the cross section is larger than that for the
$TTM$ model with $\sin\omega \geq 0.3$. For $\varepsilon=0.05$,
$200GeV\leq m_{\pi_{t}}\leq600GeV$, and $70GeV\leq E_{e}\leq200GeV$,
the value of the cross section $\sigma_{2}$ is in the range of
$5\times10^{-2}\sim 5.8\times10^{2}fb$.

Reference [21] has shown that, in most of the parameter space of the $TC2$ model, the charged top-pions $\pi_{t}^{\pm}$
mainly decay to the modes $tb$ and $cb$ with the branching ratio
about $70\%$ and $30\%$, respectively. For the decay mode $tb$,
photoproduction of the charged top-pions at the $LHeC$ would give
rise to the $t\bar{t}b$ final state, which is same as that in the
$TTM$ model. For the decay channel $\pi_{t}^{\pm}\rightarrow bc$
induced by the flavor changing interactions [17], it would give rise
to the $tbc$ final state, which induces the $l^{\pm}+bb+jet+\;
{\not\!\!E}\;$ and $bb+3jets$ signature for $W^{\pm}$ leptonic
decay, $W^{\pm}\rightarrow l\nu$ and $W^{\pm}$ hadronic decay,
$W^{\pm}\rightarrow qq'$, respectively. The number of the
$l^{\pm}+bb+jet+\; {\not\!\!E}\;$ events are shown in $Fig.5$ as
a function of the mass parameter $m_{\pi_{t}}$ for
 $E_{e}=150GeV$. one can see from this figure
that, for $200GeV\leq m_{\pi_{t}}\leq500GeV$, there will be $78\sim
3226$  $l^{\pm}+bb+jet+\; {\not\!\!E}\;$ events to be generated per
year at the $LHeC$ with ${\cal{L}}_{Lint}=50fb^{-1}$.

\vspace{0.5cm} \noindent{\bf 5. Conclusion and discussion}

\vspace{0.5cm} A common feature of the new physics model belonging
to the topcolor scenario is the prediction of the charged
top-pions in the low energy spectrum, regardless of
the dynamics responsible for $EWSB$ and light quark masses. It is well known that,
for the heavy charged scalar, which is heavier than the top quark, the
subprocesses $g b\rightarrow tS^{-}$ and $g
\bar{b}\rightarrow \bar{t}S^{+}$ are one kind of important production channels at the $LHC$. At the $LHeC$,
the charged scalars can be produced via $\gamma b$ collision, which has not been
 detailed studied in the literature, as we know. In this
paper, we consider photoproduction of the charged top-pions
predicted by the $TTM$ and $TC2$ models, which proceed via the
subprocesses $\gamma b\rightarrow t\pi_{t}^{-}$ and $\gamma
\bar{b}\rightarrow \bar{t}\pi_{t}^{+}$ at the $LHeC$.

Our numerical results show that, in the frameworks of both the $TTM$
model and the $TC2$ model, the charged top-pions $\pi_{t}^{\pm}$ can
be abundantly produced via $\gamma b$ collision, as long as they are
not too heavy. In the $TC2$ model, the production cross section is
only sensitive to the model-dependent parameter $m_{\pi_{t}}$, while
it is sensitive to the free parameters $\sin\omega$ and
$m_{\pi_{t}}$ in the $TTM$ model. For $\sin\omega<0.3$, the cross
section of the process $ep\rightarrow t\pi_{t}^{-}+X$ is larger than
that in the $TC2$ model. For example, for $\sin\omega=0.2$,
$E_{e}=150GeV$ and $200GeV\leq m_{\pi_{t}}\leq600GeV$, the
value of $\sigma_{1}(t\pi_{t}^{-})$ for the process $ep\rightarrow
t\pi_{t}^{-}+X$ is in the range of $7.3\sim995fb$, while its value is
in the range of $5\times10^{-2}\sim 5.8\times10^{2}fb$ in most of the parameter space of the $TC2$ model.

The minimal supersymmetric standard model $(MSSM)$ [22] also
predicts the existence of the charged scalars, which can also be
produced via the subprocesses $\gamma b\rightarrow tS^{-}$ and
$\gamma\bar{b}\rightarrow \bar{t}S^{+}$ at the $LHeC$. For the type II two-Higgs doublet models
($2HDMs$) [23], the tree-level production cross sections are proportional to
$(m_{t}^{2}\cot^{2}\beta + m_{b}^{2}\tan^{2}\beta)$, while proportional to
$\cot^{2}\beta$ for the type I $2HDMs$. Thus, photoproduction cross sections
of the charged scalars predicted by the $MSSM$ are larger or smaller than those
of the $TTM$ model or the $TC2$ model, which depend on the parameter $\tan\beta$.

\section*{Acknowledgments} \hspace{5mm}This work was
supported in part by the National Natural Science Foundation of
China under Grants No.10975067, the Specialized Research Fund for
the Doctoral Program of Higher Education (SRFDP) (No.200801650002),
the Natural Science Foundation of the Liaoning Scientific Committee
(No. 201102114), and Foundation of Liaoning Educational Committee (No. LT2011015).


\vspace{1.0cm}


\begin{thebibliography}{99}

\bibitem{y1} ATLAS collaboration, arXiv:1202.1408 [hep-ex]; arXiv:1202.1414 [hep-ex]; CMS collaboration, arXiv:1202.1416 [hep-ex];
 arXiv:1202.1487 [hep-ex]; arXiv:1202.1488 [hep-ex];  arXiv:1202.1489 [hep-ex].


\bibitem{y2} G. Cvetic, {\em Rev. Mod. Phys.} {\bf71}, 513(1999);
             C. T. Hill and E. H. Simmons, {\em Phys. Rept}. {\bf381}, 235(2003); [Erratum-ibid, {\bf390}, 553(2004)].

\bibitem{y3}C. T. Hill, {\em Phys. Lett. B} {\bf345}, 483(1995);
            K. D. Lane and E. Eichten, {\em Phys. Lett. B} {\bf352}, 382(1995);
            K. D. Lane, {\em Phys. Lett. B} {\bf433}, 96(1998).

\bibitem{y4}C. Csaki, C. Grojean, H. Murayama, L. Pilo, J. Terning, {\em Phys. Rev. D} {\bf69}, 055006(2004).

\bibitem{y5}R. S. Chivukula, N. D. Christensen, B. Coleppa, and E. H. Simmons, {\em Phys. Rev. D} {\bf80}, 035011(2009).

\bibitem{y6}R. S. Chivukula, E. H. Simmons, B. Coleppa, H. E. Logan, A. Martin, {\em Phys. Rev.D} {\bf83,} 055013(2011).

\bibitem{y7} Chong-Xing Yue, Hong-Jie Zong, Shun-Zhi Wang, {\em Phys. Lett. B} {\bf575}, 25(2003);
             Guo-Li Liu, {\em Phys. Rev. D} {\bf82}, 115032(2010).

\bibitem{y8} R. S. Chivukula {\em et al, Phys. Rev. D} {\bf74}, 075011(2006) .

\bibitem{y9}The LHeC web page, {\em http://www.lhec.org.uk.}

\bibitem{y10} J. B. Dainton, M. Klein, P. Newman, E. Perez, and F. Willeke, {\em JINST} {\bf1}, P10001(2006);
             P. Newman, {\em Nucl. Phys. Proc. Suppl}. {\bf191}, 307(2009);
             A. N. Akay, H. Karadeniz, S. Sultansoy, {\em Int. J. Mod. Phys. A} {\bf25}, 4589(2010).

\bibitem{y11} S. Sultansay, {\em Eur. Phys. J. C} {\bf33,} S1064(2004), and references therein.

\bibitem{y12} S. I. Alekhin et al., {\em Int. J. Mod. Phys. A} {\bf6}, 21(1991);
              A. K. Ciftci et al., {\em  Nucl. Instrum. Meth. A} {\bf365}, 317(1995);
              A. K. Ciftci, S. Sultansoy, O. Yavas, {\em Nucl. Instrum. Meth. A} {\bf472}, 72(2001);
              H. Aksakal et al., {\em  Nucl. Instrum. Meth. A} {\bf576}, 287(2007).

\bibitem{y13}  I. F. Ginzburg, G. L. Kotkin, V. G. Serbo, V. I. Telnov, {\em Nucl. Instrum. Meth}. {\bf205}, 47(1983);
               I. F. Ginzburg, et al., {\em Nucl. Instrum. Meth., A} {\bf219}, 5(1984);
               V. I. Telnov, {\em Nucl. Instrum. Meth., A} {\bf294}, 72(1990).
\bibitem{y14}J. Pumplin et al. (CTEQ Collaboration), {\em JHEP} {\bf0602}, 032(2006).

\bibitem{y15} K. Nakamura et al. [Particle Data Group], {\em J. Phys. G} {\bf37}, 075021(2010).

\bibitem{y16}R. S. Chivukula et al., {\em Phys. Rev. D} {\bf84}, 095022(2011).

\bibitem{y17}G. Burdman, {\em Phys. Rev. Lett}. {\bf83}, 2888(1999);
             H.-J. He, C.-P. Yuan, {\em Phys. Rev. Lett.} {\bf83}, 28(1999);
             H.-J. He, S. Kanemura, C.-P. Yuan, {\em Phys. Rev. Lett.} {\bf89}, 101803(2002).

\bibitem{y18}C. T. Hill, {\em Phys. Lett. B} {\bf266}, 419(1991).

\bibitem{y19}B. Balaji, {\em Phys. Lett. B} {\bf393}, 89(1997).

\bibitem{y20} G. Burdman and D. Kominis, {\em Phys. Lett. B} {\bf403}, 101(1997);
              W. Loinaz and T. Takeuchi, {\em Phys. Rev. D} {\bf60}, 015005(1999);
              C. T. Hill and X. Zhang, {\em Phys. Rev. D} {\bf51}, 3563(1995);
              Chong-Xing Yue, Yu-Ping  Kuang, Xue-Lei Wang and Wei-Bin Li, {\em Phys. Rev. D} {\bf62}, 055005(2000).


\bibitem{y21}Xue Lei Wang, Wen Na Xu, Lin Lin  Du, Comn. Theor. {\em Phys.} {\bf41}, 737(2004).

\bibitem{y22}  For reviews on the MSSM, see: P. Fayet and S. Ferrara, {\em Phys. Rep}. {\bf32}, 249(1977);
               H. P. Nilles, Phys. Rep. {\bf110}, 1(1984); R. Barbieri, Riv. Nuovo Cim. {\bf11N4}, 1(1988);
               J. Bagger, {\em Lectures at TASI-95}, hep-ph/{\bf9604232}.


\bibitem{y23} P. M. Ferreira, L. Lavoura, M. N. Rebelo, M. Sher and J. P. Silva,  \emph{arXiv:} \textbf{1106.0034} [hep-ph].

\end{thebibliography}
\end{document}